%% file: optimize_main.tex
\documentclass[9pt]{elsarticle}  

\usepackage{latexsym,amssymb,amsmath,amsfonts,dsfont,pifont}
\usepackage{fullpage,verbatim}
\usepackage{xcolor,graphicx,subfigure}
\usepackage{subfigure}

\begin{document}
\begin{frontmatter}

\title{Three Schemes for Wireless Coded Broadcast to Heterogeneous Users}

\author[rvt]{Yao Li}
\ead{yaoli@winlab.rutgers.edu}
\author[focal]{Emina Soljanin}
\ead{emina@research.bell-labs.com}
\author[rvt]{Predrag Spasojevi\'c}
\ead{spasojev@winlab.rutgers.edu}

\address[rvt]{WINLAB, Department of Electrical and Computer Engineering, Rutgers, the State University of New Jersey, 671 US 1, North Brunswick,
NJ 08902, U.S.A. }
\address[focal]{Bell Labs, Alcatel-Lucent, 600 Mountain Avenue, Murray Hill, NJ 07974 U.S.A.}

\begin{abstract}
We study and compare three coded schemes for single-server wireless broadcast of multiple description coded content
to heterogeneous users. The users (sink nodes) demand different number of descriptions over links with different packet
loss rates. The three coded schemes are based on the LT codes, growth codes, and randomized chunked codes. The schemes are
compared on the basis of the total number of transmissions required to deliver the demands of all users, which we refer
to as the server (source) delivery time. We design the degree distributions of LT codes by solving suitably defined linear
optimization problems, and numerically characterize the achievable delivery time for different coding schemes. We find that
including a systematic phase (uncoded transmission) is significantly beneficial for scenarios with low demands, and that
coding is necessary for efficiently delivering high demands. Different demand and error rate scenarios may require very
different coding schemes. Growth codes and chunked codes do not perform as well as optimized LT codes in the heterogeneous communication scenario.

\end{abstract}

\begin{keyword}
LT codes \sep chunked codes \sep growth codes \sep wireless broadcast
\end{keyword}

\end{frontmatter}

\newtheorem{thm}{Theorem}
\newtheorem{claim}[thm]{Claim}
\newtheorem{corollary}[thm]{Corollary}
\newproof{pf}{Proof}
\newdefinition{note}{Note}
\newdefinition{defn}{Definition}

\input{1intro.tex}
\input{2model.tex}
\input{3schemes.tex}

\input{4performance.tex}
\input{5conclusion.tex}

\bibliographystyle{elsarticle-num}
\bibliography{../ftrefs}

\input{appendices.tex}

\end{document}

%% file: 1intro.tex
\section{Introduction}
\subsection{Motivation}
In the past decade, the development of wireless networks has provided a fertile soil for popularization of portable digital devices, and the digital distribution of bulk media contents. This, in return, is stimulating new leaps in wireless communication technology. Today, devices retrieving digital video content in the air vary from 1080p HDTV sets to smartphones with 480$\times$320-pixel screen resolution. Under assorted restraints in hardware, power, location, and mobility, these devices experience diverse link quality, differ in computing capability needed to retrieve information from received data streams, and request information of varied granularity. Consequently, a transmission scheme designed for one type of users may not be as suitable for another even if they demand the same content.

 Today's technology implements a straightforward solution of separate transmissions to individual users, that is, multiple unicasts. Nonetheless, the key question is whether it is possible to deliver all users' demands with fewer data streams and less traffic. Especially in transmitting bulk data through wireless channels or over other shared media, reducing the amount of traffic is vital for reducing collision/interference, which in turn will also positively affect the quality of the channel in use. An additional concern in the environment conscious
 world is, of course, energy.

There is no surprise then that the problem of delivering more efficient service to a heterogeneous user community has attracted a great deal of both technical and academic interest. On the source coding side, layered coding (e.g.\cite{layered_Khansari94,layeredGallant}) and multiple description coding \cite{MDC_Goyal01} have been widely studied as solutions to providing rate scalability. In particular, with multiple description coding, a user is able to reconstruct a lower-quality version of the content upon receiving any one of the descriptions,
 and is able to improve the reconstruction quality upon receiving any additional description. Thus, the users' content reconstruction quality is commensurate with the quality of their connection. On the channel coding side,
 rateless codes \cite{luby,amin}, or fountain codes, can generate a potentially infinite stream of coded symbols that can be
 optimized simultaneously for different channel erasure rates as long as the users have uniform demands.


In this work, we explore an achievable efficiency of serving users having heterogeneous demands while using a single broadcast stream.
Whearas the information theoretical aspect of the problem is of interest and under investigation (see for example \cite{laeISIT12} and references therein), we focus on three practical coding schemes and explore their suitability for the described communications scenario.
Two important features that make codes suitable for such scenarios are (1) the ability to support partial data recovery and (2) the
ability to efficiently adapt to different channel conditions. Based on these desirable features, we chose to
investigate three candidates: LT\cite{luby}, growth\cite{growth_sigcomm06}, and chunked\cite{maymounkov_chunked,generation_ITtrans} codes. In this paper, we are particularly interested in the {\it total number of source transmissions}
needed to deliver the demands of all users. This quantity determines the amount of required communication resources, and also translates to the amount of time required for delivery. In the streaming of temporally-segmented multimedia content that is delay-constrained, it is important for the source to finish transmitting a segment as soon as possible so as to proceed to the next one. Some performance measures of interest are addressed in \cite{AllertonYE}.




\subsection{Main Results}
 We compare three coding schemes and their variations for broadcasting to heterogeneous users: users suffer different packet loss rates and demand different amounts of data. The coding schemes discussed include:
\begin{enumerate}
\item optimized LT codes, with or without a systematic phase, that is, one round of transmission of the original uncoded packets;
\item growth codes; and
\item chunked codes.
\end{enumerate}
We also compare these schemes to a reference scheme for the heterogeneous scenario based on time-shared broadcast of degraded message sets\cite{degraded}. We find that including a systematic phase is significantly beneficial towards delivering lower demands, but that coding is necessary for delivering higher demands. Different user demographics result in the suitability of very different coding schemes. Growth codes and chunked codes are not as suitable to this communication scenario as are optimized LT codes.

\subsection{Organization}
 The rest of the paper is organized as follows. Section \ref{sec:model} introduces the model of wireless broadcasting of multiple description coded content to heterogeneous users.
 Section \ref{sec:ft} introduces the coding schemes of interest, and provide theoretical characterization of code performance. In particular, the LT codes are optimized both with and without a systematic phase. In Section \ref{sec:performance} we provide numerical and simulation results of the achievable server delivery time, and discuss the suitability of these coded schemes for broadcast to heterogeneous users. The last section concludes.


%% file: 2model.tex
\section{System Model}\label{sec:model}
\begin{figure}[hbt]
\centering
\includegraphics[scale=0.5,trim=30 120 0 80]{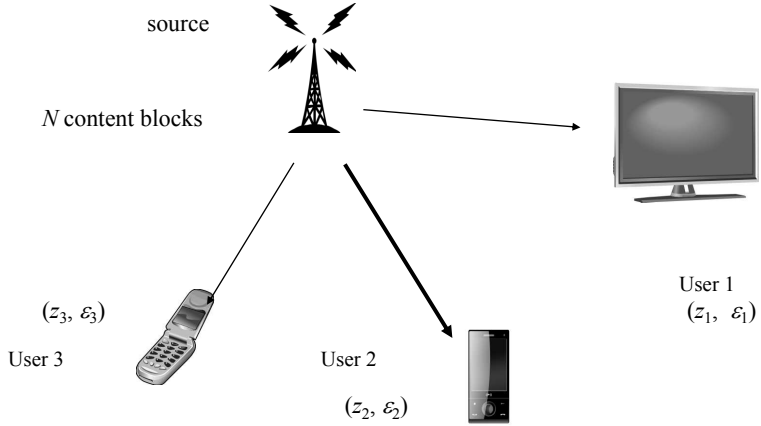}\\  
\caption{Network model}\label{fig:model}
\end{figure}

Consider a wireless single-hop broadcast network consisting of a source (server) node holding content for distribution, and $l$ sink (user) nodes waiting to retrieve the content from the broadcast stream aired by the source, as shown in Figure \ref{fig:model}.
Suppose the content held at the source node is multiple description coded into $N$ descriptions, using, for example, one of the coding schemes described in \cite{chou_layeredMDC} or in \cite{seyISIT11}, and each description is packaged into one packet for transmission. Each packet is represented as a binary vector of length $b,$ and denoted with $\xi_j$ for $j=1,2,\dots,N$.
A low-quality version of the content can be reconstructed once a user is able to recover any description.
The reconstruction quality improves progressively by recovering additional descriptions, and depends solely on the number of descriptions recovered, irrelevant of the particular recovered collection.

The source broadcasts one packet per unit time to all the sink nodes in the system. However, the sinks are connected to the source by lossy links, and at every sink, a packet either arrives at the sink intact or is entirely lost. Such an assumption is practical if we only consider data streams at the network level or higher.
Under the multiple description coding assumption, a sink node can choose to demand a smaller number of descriptions rather than wait to collect all $N$. This may not only shorten its own waiting time but also reduce the system burden. The demand of a sink node is characterized by the number of descriptions it needs to reconstruct the content within the desired distortion constraint.

Sink nodes are characterized by parameter pairs $(z_i,\epsilon_i)$, $i=1,2,\dots,l$, describing their demands and link qualities:
\begin{itemize}
\item $z_i\in\{1/N,2/N,\dots,1\}$ is the fraction of content demanded by user $i$, that is, each node demands $z_iN$ distinct descriptions.
\item $\epsilon_i$ is the packet erasure (loss) rate on each link from the source to user $i$. A packet transmitted by the source fails to reach user $i$ with probability $\epsilon_i$. All links are assumed to be memoryless.
\end{itemize}



We further assume no feedback from the sinks to the source except for the initial requests
to register the demands and the final confirmation of demand fulfillment. We also assume that the packet erasure rates $\epsilon_i$s on the links are known to the server.
In wireless broadcast, feedback from multiple nodes increases the chance of collision and compromises the point-to-point channel quality. Hence, it is desirable to keep the feedback at a minimum.

\begin{defn}(General Delivery Time)
The {\it delivery time} $T_i$ of a sink is a random variable defined as the number of packets transmitted from the server until
the fulfillment of the demand $z_i$ of user $i$. The {\it server delivery time} $T=\max_{i=1}^l{T_i}$ is the number of packets transmitted from the server required to fulfill the demands of all the users.
\end{defn}

Since all users retrieve information from the same broadcast stream through channels of random erasures, $T_i$'s and $T$ are
 dependent random variables. Nevertheless, even their marginal probability distributions are not easy to characterize. Instead, in this paper,
 we normalize the delivery time by the total number of content packets $N$, and restrict our attention to either the expectation
 $t_i=E[T_i]/N$ or the asymptotics $t_i=\lim_{N\rightarrow\infty}T_i/N$,
 and study $\max_{i=1}^{l}t_i$ as the (normalized) server delivery time. When $t_i=E[T_i]/N$, $\max_{i=1}^{l}t_i$ is a lower bound for $E[T]/N$,
  the normalized expected delivery time. For brevity we further abuse
 our notation and terminology and specialize the definition of delivery time to these lower bounds. The specialization of a definition will be stated where appropriate.
In the rest of this paper, we describe and analyze three randomized coding schemes
for broadcast in a heterogeneous setting, optimize if appropriate, compare their performance, and discuss their suitability to our end.

Note that with multiple unicasts, the total number of packets transmitted would be (in the mean) at least $\sum_{i=1}^l\frac{z_i}{1-\epsilon_i}.$
We will show in our results that it is possible to do better using broadcast.


%% file: 3schemes.tex
\section{Three Coding Schemes for Broadcast to Heterogeneous Users} \label{sec:ft}
In this section, we describe three coding schemes and characterize their delivery time performance for broadcast to heterogeneous users.
All the coding schemes studied in this paper are applied at the packet level. Each coded packet is obtained as a random linear combination of
the original packets, but either the number or the range of the packets involved in each linear combination is restricted.

\subsection{LT Coded Broadcast}\label{sec:ft_ft}
LT codes were first proposed by Michael Luby in \cite{luby} as a class of codes
designed to overcome variations in channel quality via ratelessness. Unlike their rateless predecessors, the ordinary random linear codes,
LT codes can be decoded by simple suboptimal decoders with little loss in performance.
In \cite{sanghavi}, Sujay Sanghavi pointed out the inadequacy of the original LT codes for partial data recovery, and demonstrated the possibility of redesigning the LT codes. His focus was on scenarios with users having uniform demands. Here we extend the code optimization framework of \cite{sanghavi}
to adapt the code to the heterogeneous scenario in which both the channel conditions and the demands vary among users.

\subsubsection{Overview of LT Encoding and Decoding}\label{sec:lt_overview}

The LT codes, as defined in Luby's original paper \cite{luby}, are binary. An LT encoder outputs a potentially infinite stream of random linear combinations of the $N$ original content packets, and decodes with a belief-propagation decoder. Each coded packet is generated independently as follows: first, pick a {\it degree} $d$ according to a distribution defined by its moment generating function
\begin{equation}\label{eq:mgf}
P^{(N)}(x)=p_1x+p_2x^2+\ldots+p_Nx^N,
\end{equation}
where $p_j\triangleq\mathrm{Prob}[d=j]$; then, select with equal probability one of the ${N \choose d}$ possible combinations of $d$ distinct packets from the $N$ content packets, and form a coded packet of degree $d$ by linearly combining the chosen packets over $\mathrm{GF}(2)$. The linear combination is represented by a coding vector $c=[c_1,c_2,\dots,c_N]^T$, a binary column vector of length $N$ with exactly $d$ non-zero components corresponding to the chosen packets. The generated coded packet is given by $\sum_{j=1}^{N} c_j\xi_j$ (recall $\xi_j$s are the binary vector representations of the content packets as defined in Section \ref{sec:model}) and it is of the same size $b$ as the original content packets. The $N$-bit coding vector $c$ can be embedded into the coded packet and transmitted to the user. In wireless communications where data packets are generally of at most a few kilobytes' size, if $N\sim$1K, it is costly to add an extra $N$ bits. Instead, the source can embed into the packet the degree of the coding vector and the seed of a pseudo-random number generator that is in shared knowledge of both the source and all the sink nodes, to allow the sink node to regenerate the coding vector locally.

Decoding is done by a belief propagation decoder\cite{luby,amin}. The decoder maintains a set called the {\it ripple}. The initial ripple is composed of all received coded packets of degree-1. In each decoding step, one coded packet in the ripple is processed by removing it from the ripple, substituting it into all the coded packets it participates in, which reduces the degrees of these coded packets. The coded packets whose degrees are reduced to 1 are added to the ripple, and the decoder continue to process another coded packet in the ripple. The decoding
process halts when the ripple becomes empty.

\subsubsection{Characterization of the Recoverable Fraction}
Each user $i$ needs to recover $z_iN$ content packets from the received coded packets.
\begin{defn}(Recoverable Fraction)
 A fraction $z$ is said to be {\it recoverable} by the belief propagation decoder if the ripple size stays positive (at least) until $zN$ packets get decoded.
\end{defn}


Based on Theorem 2 in \cite{2nd_moment}, Corollary \ref{thm:ft_randchannel} gives the expected ripple size as a function of the recovered fraction of the content packets. We restate the part of Theorem 2 in
\cite{2nd_moment} that concerns the expected size of the ripple as Theorem \ref{thm:ripplesize}.

Assume $w\cdot N$ coded packets have been collected and fed into an LT decoder, for some positive constant $w$. Let
$u\cdot N$ be the number of decoded content packets, for a
constant $u\in[0,1)$. Let $r^{(N)}(u)$ be the expected size of the
ripple, normalized by $N$.

\begin{thm}(Maatouk and Shokrollahi~\cite[Thm.~2]{2nd_moment})\label{thm:ripplesize}
If an LT code of $N$ content packets has degree distribution
specified by the moment generating function $P^{(N)}(x)$ (see (\ref{eq:mgf})), then
\begin{equation}\label{eq:ripplesize_N}
r^{(N)}(u)=wu\Bigl(P^{(N)\prime}(1-u)+\frac{1}{w}\ln
u\Bigr)+\mathcal{O}\Bigl(\frac{1}{N}\Bigr),
\end{equation}
where $P^{(N)\prime}(x)$ stands for the first derivative of
$P^{(N)}(x)$ with respect to $x$.
\end{thm}

We further extend Theorem \ref{thm:ripplesize} to the case where the number
of collected coded packets is random to accommodate random packets losses over the channel.
Assume the number of collected coded packets is now $W\cdot N$, where $W$ is a random variable with mean $v.$
Denote the normalized expected ripple size as $N\rightarrow\infty$ as $r_W(u)$.
Assuming that $P^{(N)}(x)$ converges to $P(x)=\sum_{i\ge1}p_ix^i$ as $N\rightarrow\infty,$ then the following holds:

\begin{corollary}\label{thm:ft_randchannel}
\begin{equation}
r_W(u)=u\Bigl(v P^{\prime}(1-u)+\ln u\Bigr).
\label{eqn:randomchannel}
\end{equation}
\end{corollary}
\begin{pf}
Although Theorem \ref{thm:ripplesize} (Theorem 2 in \cite{2nd_moment}) is stated for the case
where the number of coded packets collected by the sink node is more
than the total number of content blocks, i.e., $w>1$,
its proof suggests that the theorem also holds for any constant
$w<1$.

Take $N\rightarrow\infty$ on both sides of \ref{eq:ripplesize_N},
we have
\begin{equation}
r(u)=\lim_{N\rightarrow\infty}r^{(N)}(u)= u\bigl(w P'(1-u)+\ln
u\bigr). \label{aymp_ripple}
\end{equation}

Replace $w$ in (\ref{aymp_ripple}) with $W.$ Due to the linearity of the expected ripple size in $W$ for
given $u$ and $P$,
\begin{eqnarray*}
r_W(u) &=&E\left[u\Bigl(WP^{\prime}(1-u)+\ln
u\Bigr)\right]
=u\Bigl(v P^{\prime}(1-u)+\ln u\Bigr)
\end{eqnarray*}
\end{pf}



If we use the expected value as a rough estimate of the ripple size during the decoding process, we should have $r_W(u)>0$ for $u\in(1-z,1].$
Applying \eqref{eqn:randomchannel}, we have
\begin{equation}
v P'(1-u)+\ln u>0, \quad\forall u\in(1-z,1],
\label{eqn:rand_recovery_condition}
\end{equation}
We next use \eqref{eqn:rand_recovery_condition} as a constraint to
formulate an optimization problem for LT code degree distribution
design to minimize the number of transmissions required to meet all sink demands.

\subsubsection{Minimizing Server Delivery Time by Degree Distribution Design}\label{sec:ft_nonsys_opt}
We are concerned with the delivery time of LT coded broadcast in the asymptotic regime when $N\rightarrow\infty$.
\begin{defn}(LT Delivery Time)\label{def:lt}
For the LT coded scheme, the (normalized) {\it delivery time} $t_i$ is defined as the ratio of the number of transmissions required to fulfill the demand $z_i$ of user $i$ to the the total number $N$ of content packets, as $N\rightarrow\infty$. The (normalized) {\it server delivery time} is taken as $t_0=\max_{i=1}^{l}\{t_i\}$.
\end{defn}

The normalized number of coded packets user $i$ receives by its delivery time $t_i$ over a channel with the
packet erasure rate $\epsilon_i$ is on average $t_i(1-\epsilon_i)$.
Let $x=1-u$ in
(\ref{eqn:rand_recovery_condition}) and $\omega=t_i(1-\epsilon_i)$; we have
\begin{equation}
(1-\epsilon_i)t_iP'(x)+\ln(1-x)>0, \quad\forall x\in[0,z_i),
\label{eqn:rec_dec}
\end{equation}
and consequently,
\begin{equation}\label{eq:lt_find_ti}
t_i=\inf\{\tau:(1-\epsilon_i)\tau P'(x)+\ln(1-x)>0, \forall x\in[0,z_i)\}.
\end{equation}
On the other hand, the recovered fraction by the user on a link of loss rate $\epsilon$ as a function of the number $t$ of transmitted packets (normalized by $N$) is
\begin{equation}\label{eq:lt_findz}
z(t,\epsilon)=\sup\{\zeta:(1-\epsilon)tP'(x)+\ln(1-x)>0, \forall x\in[0,\zeta)\},
\end{equation}
i.e., the recoverable fraction is the maximum $\zeta$ such that the expected ripple size stays positive in the whole range of $[0,\zeta].$

We now have all elements to state an optimization problem for minimize the server delivery time $t$.
Using (\ref{eqn:rec_dec}), the optimization problem is expressed as follows:

 \begin{align}
 \mbox{min.}_{P,t_1,\dots,t_l}\quad& t_0=\max_{i}\ t_i    \label{eq:opt_gen_lat_fair}\\
 \mbox{s.t.}\quad&t_i(1-\epsilon_i)P'(x)+\ln(1-x) > 0, \quad 0\le x\le z_i,\quad\mbox{for }i=1,2,\dots,l; \notag\\
 &P(1)=1,\notag
 \end{align}
 or equivalently, 
\begin{eqnarray}
&\mbox{min.}_{P,t_0}& t_0    \label{eq:opt_gen_lat_fair0}\\
&\mbox{s.t.}&t_0(1-\epsilon_i)P'(x)+\ln(1-x) > 0,\quad 0\le x\le z_i, \quad\mbox{for }i=1,2,\dots,l.\notag\\
&&P(1)=1.\notag
\end{eqnarray}

\begin{claim}
There must exist an optimal solution to Problem
\eqref{eq:opt_gen_lat_fair0} with a
polynomial $P(x)$ of degree no higher than
$d_{\max}=\lceil\frac{1}{1-\max_i\{z_i\}}\rceil-1$.
\end{claim}
\begin{pf}
This claim is provable with an argument similar to Lemma 2 of \cite{sanghavi}.
Suppose $(t^*, P^*)$ is an optimal solution of Problem \eqref{eq:opt_gen_lat_fair0}. construct $\bar{P}$ such that
$\bar p_{j}=p^*_{j}$ for $j=1,2,\dots,d_{\max}-1$, and $\bar p_{d_{\max}}=\sum_{j\ge d_{\max}}p^*_{j}.$ Then $\bar P(x)$ still represents a degree distribution, and meanwhile,
\begin{align*}
\bar P'(x)-P^{*\prime}(x)&=\sum_{j= 1}^{d_{\max}}j\bar p_jx^{j-1}- \sum_{j\ge 1}jp^*_jx^{j-1}\\
&=\sum_{j=1}^{d_{\max}-1}jp^*_jx^{j-1} + d_{\max}\sum_{j\ge d_{\max}}p^*_jx^{d_{\max}-1} - \sum_{j= 1}^{d_{\max}-1}jp^*_jx^{j-1}-\sum_{j\ge d_{\max}}jp^*_jx^{j-1}\\
&=\sum_{j\ge d_{\max}}[d_{\max}p^*_jx^{d_{\max}-1}- jp^*_jx^{j-1}]\\
&=\sum_{j\ge d_{\max}+1}p^*_jx^{d_{\max}-1}[d_{\max}- jx^{j-d_{\max}}]\\
&\ge \sum_{j\ge d_{\max}+1}p^*_jx^{d_{\max}-1}[d_{\max}- (d_{\max}+1)x].\\
\end{align*}
As long as $x\le \frac{d_{\max}}{d_{\max}+1},$ i.e., $d_{\max}\ge \frac{x}{1-x}=\frac{1}{1-x}-1,$ $\bar P'(x)\ge P^{*\prime}(x).$ Thus, set $d_{\max}=\lceil\frac{1}{1-\max_i{z_i}}\rceil-1\ge \lceil\frac{1}{1-z_i}\rceil-1\ge\frac{1}{1-z_i}-1,$ we have for $i=1,2,\dots,l,$
\[t^*(1-\epsilon_i)\bar P'(x)+\ln(1-x)\ge t^*(1-\epsilon_i) P^{*\prime}(x)+\ln(1-x)>0,\quad 0\le x\le z_i,\]
and hence $(t^*,\bar P)$ is also a feasible and optimal solution of Problem \eqref{eq:opt_gen_lat_fair0} with optimal value $t^*,$ and the highest degree of $\bar P$ is no more than $d_{\max}=\lceil\frac{1}{1-\max_i\{z_i\}}\rceil-1.$
\end{pf}

Thus, Problem (\ref{eq:opt_gen_lat_fair0})
can readily be converted into a linear programming problem by the method proposed in
\cite{sanghavi}. For $j=1,2,\dots,d_{\max}$, let $a_j=tp_j$, and Problem (\ref{eq:opt_gen_lat_fair0}) becomes
\begin{eqnarray}
&\mbox{min.}_{a_1,\dots,a_{d_{\max}}}& \sum_{j=1}^{d_{\max}}a_j    \label{eq:opt_gen_lat_fair_linear}\\
&\mbox{s.t.}& \sum_{j=1}^{d_{\max}}ja_jx^{j-1}  > -\frac{\ln(1-x)}{1-\epsilon_i},\quad 0\le x\le z_i, \quad\mbox{for }i=1,2,\dots,l.\notag\\
&&a_j\ge0, j=1,2,\dots,d_{\max}.\notag
\end{eqnarray}

To solve the linear programming problem (\ref{eq:opt_gen_lat_fair_linear})
numerically, the constraints defined on a continuous interval of parameter $x$ are written out by evaluating $x$ at
discrete points within the interval. Lower bounds for the minimum value of \eqref{eq:opt_gen_lat_fair_linear}, and \eqref{eq:opt_gen_lat_fair} can thus be obtained. (In Section \ref{sec:performance}, we interestingly observe that in a 2-user scenario, the optimal server delivery time obtained from the optimization described here is close to the delivery time achievable by using a time-sharing scheme to broadcast degraded message sets. The time-sharing scheme is described in \ref{sec:ref}.)

\subsubsection{LT Coding with a Systematic (Uncoded) Phase}\label{sec:ft_sys_opt}
We also study a variation of the LT codes that start with a systematic phase, namely, transmission of all the original uncoded content packets (systematic packets) followed by parity packets.

The formulation of the degree distribution optimization problem is essentially the same, except that \eqref{eqn:rec_dec} (describing the condition which the number of transmissions and the degree distribution must satisfy to allow the delivery of the demand of each user $i$) becomes constraint \eqref{eq:rec_dec_sys} in the following Claim \ref{thm:sys_rec}.

\begin{claim}\label{thm:sys_rec}
Suppose after the systematic phase, the  degree distribution of the parity packets transmitted subsequently follows the distribution represented by $P(x),$ as $N\rightarrow\infty.$ Then, to recover a fraction $z_i$, the normalized number of parity packets transmitted should satisfy
\begin{equation}
-\ln(\epsilon_i)+(1-\epsilon_i)(t_i-1)P'(x)+\ln(1-x)>0, \quad\forall x\in(1-\epsilon_i,z_i).
\label{eq:rec_dec_sys}
\end{equation}
\end{claim}
\begin{pf}
Please refer to the Appendix.
\end{pf}

The new optimization problem is still readily transformable into a linear programming problem.

In addition, we have
\begin{equation}\label{eq:lt_sys_find_ti}
t_i=\left\{\begin{array}{lr}\frac{z_i}{1-\epsilon_i},&z_i\le 1-\epsilon_i;\\
\inf\{\tau:-\ln(\epsilon_i)+(1-\epsilon_i)(\tau_i-1)P'(x)+\ln(1-x)>0,\forall x\in[0,z_i) \},&z_i> 1-\epsilon_i.\end{array}\right.
\end{equation}

The systematic phase delivers the demand of a user with $z_i\le 1-\epsilon_i$, which helps reduce the server delivery time.
It is similarly possible to formulate optimization problems to allow transmitting less or more than one round of systematic symbols, and find
out the tradeoff between the fraction of systematic symbols and non-systematic symbols. However, this is beyond the scope of this paper.

\subsection{Growth Codes} \label{sec:growth}
Growth codes were proposed by Kamra et al.\ in \cite{growth_sigcomm06} to improve data persistence in sensor networks in face of sensor node failure. Growth codes were not designed for our heterogeneous scenario described in Section \ref{sec:model}. However, their feature of progressive partial recovery suggests that they may be a good candidate scheme. Extensions and applications of growth codes to video streaming
has been studied in, e.g., \cite{dimakisgrowth, h.264growth}. Particularly, in \cite{dimakisgrowth}, a systematic version of growth codes with unequal protection for layer coded video content was proposed.

Growth coded packets are, as LT coded packets, binary random linear combinations of the original source packets, and can be decoded by a belief propagation decoder that is essentially identical to that of the LT codes, as described in Section \ref{sec:lt_overview}. But, unlike an LT coded stream which produces statistically identical packets, a growth coded stream starts with degree-1 coded packets and gradually move on to send coded packets of higher degrees.
The encoding scheme described in \cite{growth_sigcomm06} operates as follows. Let $R_j=\frac{jN-1}{j+1}$ for $j=0,1,2,\dots,N-1$. Let $A_j=\sum_{s=\lfloor  R_{j-1}\rfloor+1}^{\lfloor R_j\rfloor}\frac{{N\choose j}}{{s\choose j}(N-s)}$ for $j=1,2,\dots,N-1$.
Then, on a perfect erasure free channel, the source node sends $A_1$ degree-1 coded packets followed by $A_2$ degree-2 coded packets, $A_3$ degree-3 packets, and so on. Demand of size between $R_{j-1}$ and $R_j$ is expected to be fulfilled during the phase when degree-$j$ coded packets are sent.

 Growth codes are based on the design philosophy to greedily ensure the highest probability of recovering a
 new content packet upon receiving each additional coded packet. On a link with loss probability $\epsilon$, it is reasonable to scale each duration $A_j$ in which degree-$j$ coded packets are transmitted by $1/(1-\epsilon)$ so as to keep the degree distribution of the packets reaching the sink approximately the same as if the code runs on a perfect channel. There is no straightforward way to extend such design philosophy and scaling approach to broadcasting over channels of different erasure rates,
 but we can still scale $A_j$s for one of the users and see the resulting delivery time for other users and the server, and search for the scaling factor with which the server delivery time is minimized. We use the belief propagation LT decoder to decode growth codes, and use \eqref{eq:lt_find_ti} to predict the delivery time (as defined in Definition \ref{def:lt}) and recoverable fraction of the scaled version of growth codes as $N\rightarrow\infty$. We compare growth codes with the optimized LT codes in our heterogenous scenarios in Section \ref{sec:performance}.

\subsection{Chunked Codes}
\label{sec:chunked}
 Chunked coding was first proposed and studied in \cite{maymounkov_chunked}. It has also been investigated by the authors of this paper in \cite{generation_ITtrans} via a coupon collection analysis. The $N$ content packets are grouped into $n$ disjoint subsets of $h$ packets(assuming $N$ is a multiple of $h$), and these subsets are called ``chunks''. Packets are represented as vectors of symbols from $\mathrm{GF}(q)$. In each transmission, we first uniformly, randomly select a chunk, and then sample from $\mathrm{GF}^h(q)$ a coding vector uniformly at random and form a linear combination of the content packets within the selected chunk. As soon as a sink node has collected $h$ coded packets with linearly independent coding vectors generated from the same chunk, all the packets of this chunk can be decoded by performing Gaussian elimination on $\mathrm{GF}(q)$. As opposed to full network coding, with which coding vectors are chosen from $\mathrm{GF}^N(q)$, this scheme has lower computational complexity and also to some extent allows partial recovery.

From Theorem 4 of \cite{generation_ITtrans}, we know that, for a sufficiently large field size $q$, the expected number of transmissions needed to decode any $k$ of the $n$ chunks on a unicast link subject to packet loss rate of $\epsilon$ is given as
\begin{equation}\label{eq:chunk_et}
E[T(n,k,\epsilon)]=\frac{n}{1-\epsilon}\int_{0}^{\infty}\left\{\sum_{j=0}^{k-1}{n\choose
i}S_{h}^{n-j}(x)\left[e^x-S_{h}(x)\right]^j\right\}e^{-nx}dx,
\end{equation}
where
\begin{align*}
S_m(x)=&1+\frac{x}{1!}+\frac{x^2}{2!}+\dots+\frac{x^{m-1}}{(m-1)!}\quad(m\ge
1)\\
S_{\infty}(x)=& \exp(x) ~\text{and} ~ S_{0}(x)=0.
\end{align*}
The above result is based on the generalized birthday problem in \cite{Flajolet1992207}.

\begin{defn}(Chunked Codes Delivery Time)
With the chunked-coding scheme, the normalized delivery time of user $i$ is defined as $t_i=E[T(n,\lceil z_in\rceil,\epsilon_i)]/N$.
The normalized server delivery time is defined as $t_0=\max_{i=1}^lt_i$.
\end{defn}

\subsection{Lower Bound and Reference Schemes}
\subsubsection{A Lower Bound}
An obvious lower bound for the minimum server delivery time is $t=\max_{i=1}^l\{\frac{z_i}{1-\epsilon_i}\}.$

\subsubsection{Multiple Unicasts}
If instead of broadcast, the server transmits separate streams to each user, the total normalized number of transmissions required will be at least $t=\sum_{i=1}^l\frac{z_i}{1-\epsilon_i}.$

\subsubsection{Braodcast Degraded Message Sets by Timesharing}\label{sec:ref}
Here we describe a reference coding scheme. 
Without loss of generality, assume $z_0=0<z_1\le z_2\le\dots\le z_l$. Then, segment $N$ descriptions/packets into $l$ layers, with Layer $i(i=1,2,\dots,l)$ containing $L_i=(z_{i}-z_{i-1})N$ packets. Protect Layer $i$ by an erasure code of rate $R_i=1-\max\{\epsilon_i,\epsilon_{i+1}, \dots, \epsilon_l\},$ and transmit the protected layers sequentially. When $N$ goes to infinity, there exist erasure codes that allow the server to deliver Layers $1$ through $i,$ that is $z_iN$ packets, to user $i$ for all $i=1,2,\dots,l$ in $\sum_{i=1}^l\frac{L_i}{R_i}$ time. Later we will find in Section \ref{sec:performance_delivery_time} that in a 2-user scenario the server delivery time of this scheme is close to that of the optimized LT codes without a systematic phase.

%% file: 4performance.tex
\section{Performance Comparison}\label{sec:performance}
In this section we evaluate the schemes described in Section \ref{sec:ft} by numerical calculation and simulation for a 2-user scenario.
\subsection{Partial Recovery Curves}\label{sec:performance_partial}
\begin{figure}[hbt]
\centering
\subfigure[Numerical Results]{
\includegraphics[scale=0.42]{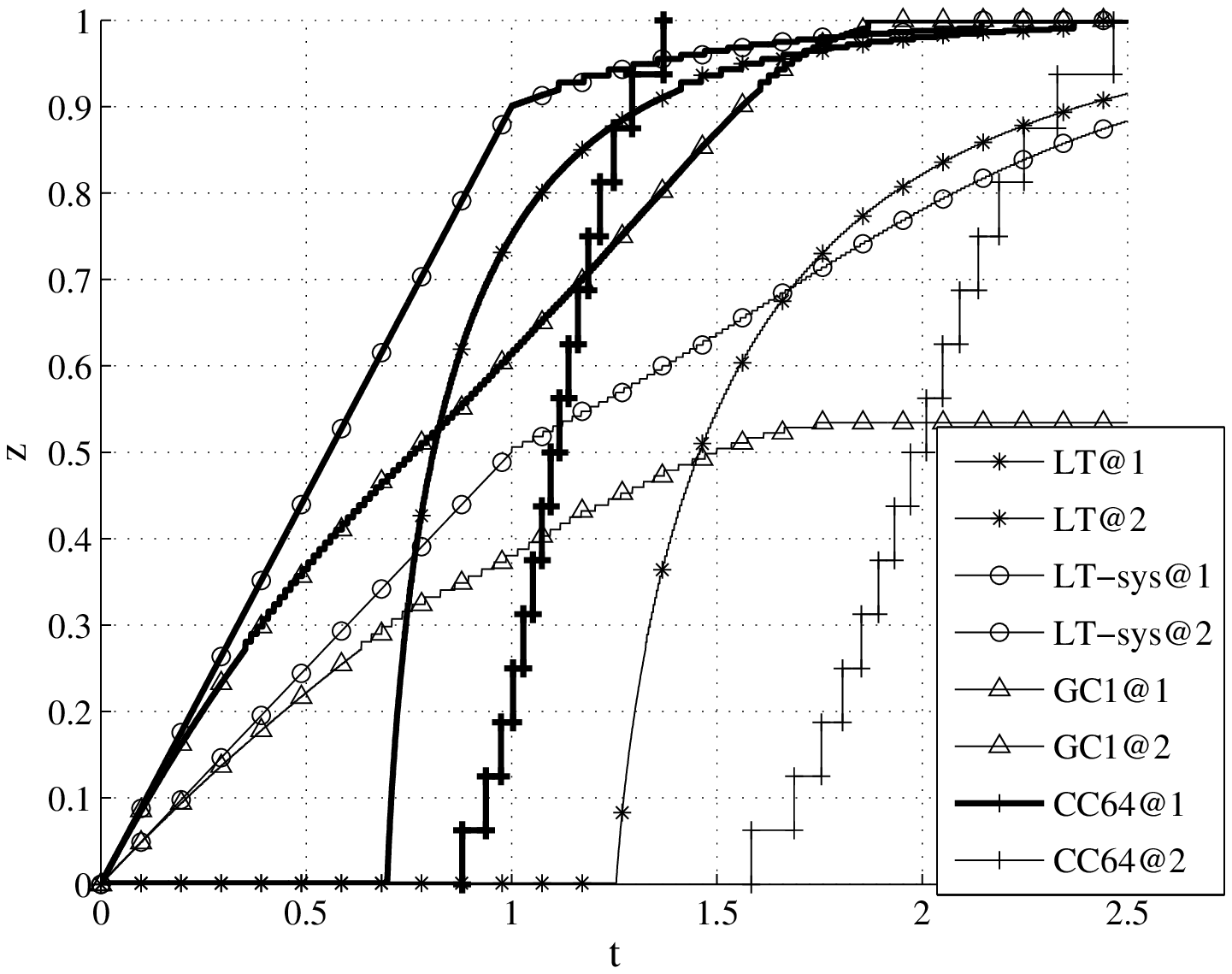}  
\label{subfig:partial_theory}
}
\subfigure[Simulation Results]{
\includegraphics[scale=0.42]{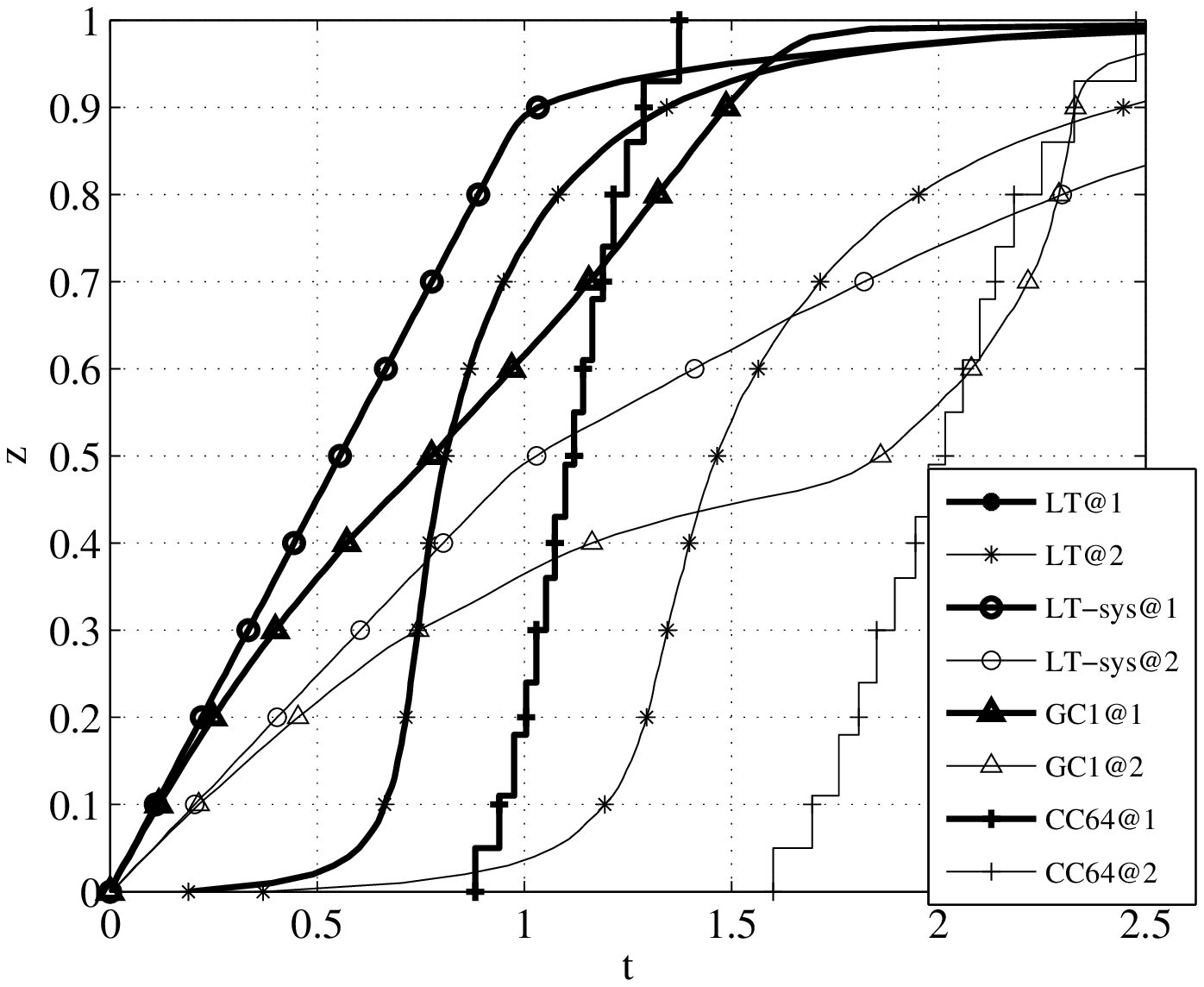} 
\label{subfig:partial_sim}
}
\caption[Optional caption for list of figures]{Evolution of recoverable fraction with time at each user in a 2-user scenario. $\epsilon_1=0.1,$ $\epsilon_2=0.5.$ $N=1024.$
Schemes: (1) LT optimized with $z_1=15/16$ and $z_2=9/16.$ (2)optimal systematic LT
(3) growth codes with a scaling factor of $\frac{1}{1-\epsilon_1}$ (4) chunked codes with $n=16$ chunks of $64$ packets.
LT(-sys): Optimized LT codes (with a systematic phase); GC1: Growth codes scaled by $\frac{1}{1-\epsilon_1};$ CC$_{64}$: Chunked codes with 64 packets per chunk.
}\label{fig:partial}
\end{figure}

We demonstrate in Figure \ref{fig:partial} the evolution of the fraction of recoverable content packets at sink nodes with the growth of the number of transmissions from the source in the 2-user broadcast scenario. The erasure rates are $\epsilon_1=0.1$ and $\epsilon_2=0.5.$ User $2$ has a worse channel. Shown are the performance curves of optimized LT codes (with and without a systematic phase), growth codes, and chunked codes at both users.
The details in obtaining the numerical results are listed below.
\begin{itemize}
\item LT codes (systematic and non-systematic)
    \begin{itemize}
    \item The degree distributions are obtained by solving the optimization problems in \ref{sec:ft_nonsys_opt} and \ref{sec:ft_sys_opt} by setting $(z_1,\epsilon_1)=(15/16,0.1)$ and $(z_2,\epsilon_2)=(9/16,0.5)$. For the non-systematic version, the optimal $P(x)=0.0195x+0.7814x^2+0.1991x^3;$ for the systematic version, the optimal $P(x)=0.7061x^2+0.2939x^3.$
    \item The delivery time is computed for given $z_1$ and $z_2$ based on \eqref{eq:lt_find_ti} in the nonsystematic case and on \eqref{eq:lt_sys_find_ti} in the systematic case. The optimal server delivery time is $1.5178$ for the nonsystematic version, and $1.2488$ in the systematic version.
    \end{itemize}
\item Growth codes
    \begin{itemize}
    \item $N=1024.$ The time spent transmitting coded packets of each degree $i$, $A_i$, is scaled by a factor of $1/(1-\epsilon_1)$, that is, the code is adapted to the channel conditions of user $1$.
    \item The recoverable fraction is computed by evaluating \eqref{eq:lt_findz} on the empirical degree distribution of received packets. Note that \eqref{eq:lt_findz} assumes $N\rightarrow\infty$.
    \end{itemize}
\item Chunked Codes
    \begin{itemize}
    \item We set the number of packets to $N=1024$, the number of chunks to $n=16,$ and thus the chunk size becomes $h=N/n=64$.
    \item We compute $E[T(n,\lceil zn\rceil,\epsilon_i)]/N$ with $E[T(n,k,\epsilon)]$ given by \eqref{eq:chunk_et}.
    \end{itemize}
\end{itemize}

Simulation results show the average time required for the user to recover a given fraction $z$ of all the packets. The average is taken over
$100$ runs, and the number of source packets $N$ is set to be $1024$ in all runs. 
For chunked codes, the finite field size $q$ is set to be $256$. It is shown that the simulation results are close to the theoretical performance prediction for the coding schemes studied, except for the difference in the latter stage of growth codes, which comes from the slight difference in the coding scheme numerically evaluated and the one simulated. In our simulations, after finishing sending $A_m$ degree-$m$ coded packets where $m=\lceil\frac{1}{\max_{i=1,2}\{1-z_i\}}\rceil$, instead of proceeding to sending packets of higher degrees, as defined in the original scheme, we allow the server to send coded packets according to a degree distribution $p_j=\frac{A_j}{\sum_{j=1}^m A_j}$ for $j=1,2,\dots,m,$ the cumulative degree distribution up to stage $A_m.$ This provides the user on the worse channel with the low-degree packets that are needed but have been lost due to a higher packet loss rate, and allows the user to proceed in packet recovery.
All the coding schemes exhibit partial recovery property to a corresponding degree, namely, the recoverable fraction gradually increases with time. With both optimized LT (without the systematic phase) and the chunked codes, there is an initial stage where no packets are recoverable.

\subsection{Delivery Time}\label{sec:performance_delivery_time}
\begin{figure}[hbt]
  \centering
  \includegraphics[scale=0.8]{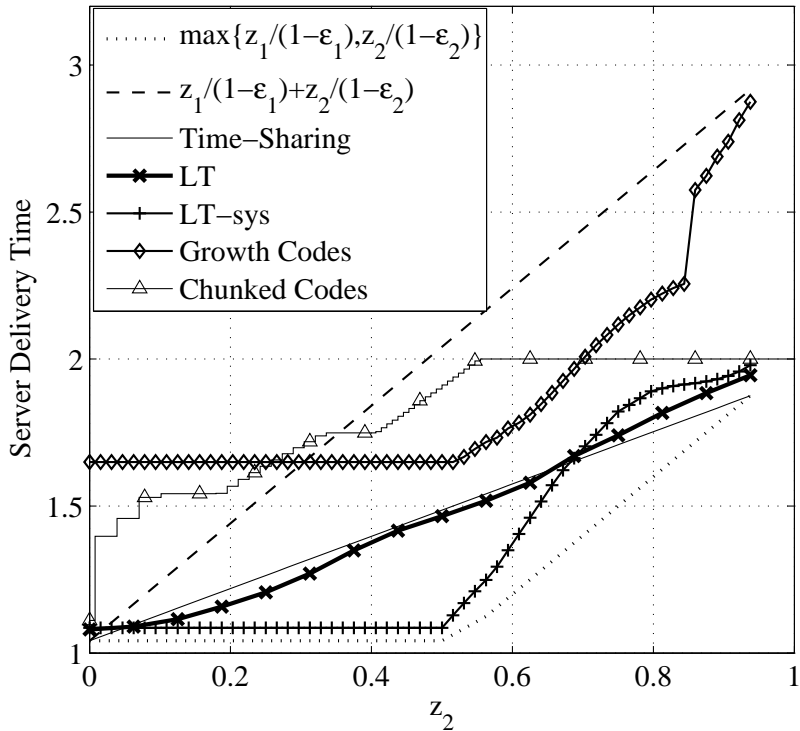}\\ 
  \caption{Comparison of the server delivery time $t$: $N=1024$ descriptions/packets ($N\rightarrow\infty$ for LT codes and growth codes), $\epsilon_1=0.1$, $\epsilon_2=0.5$, $z_1=15/16$, $z_2$ ranging from 0 to $15/16$.
  }\label{fig:delivery_ft}
\end{figure}

We continue with a study of the server delivery time performance of the coding schemes in the 2-user scenario. We keep $\epsilon_1=0.1$ and $\epsilon_2=0.5,$ set $z_1=15/16$, let $z_2$ vary from $0$ to $15/16$, and plot the server delivery time versus $z_2$. With LT codes, for each set of $(z_1,z_2)$ values, we solve for the optimal degree distribution. With growth codes, for each $z_2,$ we search all scaling factors between $[\frac{1}{1-\epsilon_1},\frac{1}{1-\epsilon_2}]$ for the one that minimizes the server delivery time. With chunked codes, we examine all power-of-2 chunk sizes between $h=2^0=1$ and $h=2^{10}=N,$ and find the chunk size that minimizes the server delivery time.
The numerical evaluation of the minimized server delivery time is plotted versus the demand $z_2$ of user 2 in Figure \ref{fig:delivery_ft}. Note that the optimal degree distribution changes as the demands vary.


The server delivery time of the reference scheme via time-shared broadcast of degraded message sets, as described in \ref{sec:ref}, is also plotted. In the two-user case discussed here, $z_1\ge z_2$ and $\epsilon_1>\epsilon_2$. Therefore, two layers are formed, the first consisting of $L_1=z_2N$ blocks and the second another $L_2=(z_1-z_2)N$ blocks, and respectively protected with erasure codes of rates $1-\epsilon_1$ and $1-\epsilon_2$. It is particularly interesting to find that in the setting demonstrated here, the reference scheme performance almost coincides with the non-systematic optimized LT coded scheme with optimized degree distribution. The reason of such a phenomenon, however, awaits further investigation and is beyond the scope of this work.

In addition, in Figure \ref{fig:delivery_ft}, we plot two other reference lines: $t=\max\{\frac{z_1}{1-\epsilon_1},\frac{z_2}{1-\epsilon_2}\} $ and $t=\frac{z_1}{1-\epsilon_1}+\frac{z_2}{1-\epsilon_2}$. The former represents an apparent lower bound for server delivery time, and the latter represents a lower bound for the total number of packet transmissions if the demands are delivered by unicasts to each individual user.

Comparing the server delivery time of systematic and nonsystematic LT codes optimized with the knowledge of channel state and user demands, we find that the systematic phase significantly shortens the server delivery time when the demand of the user on the worse channel is low. As the demand of user 2 rises, however, including a systematic phase becomes suboptimal, until the demand rises further closer to $1$, when the delivery time seems to converge to the delivery time with the nonsystematic scheme.

Growth codes and chunked codes do not have a competitive server delivery time performance compared with other schemes. When the demand of the user on the worse channel is low, using these schemes requires more transmissions to deliver the demands than using multiple unicasts. One advantage
of chunked codes, however, is that all the content packets in the same chunk are decoded simultaneously, allowing these packets
 to be dependent on each other for the reconstruction purpose, which may reduce redundancy in the source coding stage.

It is also of interest to study performance measures other than the delivery time of the server. For example, one could wish
to maximize the minimum of users' throughputs $\min_i\{z_i/t_i\}$, the ratio of the demand and the time needed to complete the demand, or to maximize the minimum channel utilization $\min_i\{z_i/((1-\epsilon_i)t_i)\}$, the ratio of the information pushed through the channel to the channel bandwidth. These criteria are of interest for acieving, e.g., fairness in the network, and optimization for these criteria can yield very different code design parameters (degree distributions). The optimization of the degree distributions for LT coded broadcast for these different criteria has been studied in our work \cite{AllertonYE} and interested readers are kindly referred to
this paper for detailed problem formulation and results. An additional type of heterogeneity can also be treated in the framework provided in Section \ref{sec:ft_ft}, namely, when some of the sink nodes cannot decode but can only recover content from degree-1 coded packets. Readers are also referred to \cite{AllertonYE} for results regarding the coexistence of nodes able and unable to decode in the system.

%% file: 5conclusion.tex
\section{Conclusion and Future Work}
We investigated the usage of three coded schemes for broadcasting multiple description coded content
in a single-hop wireless network with heterogeneous user nodes of nonuniform demand over links of diverse packet loss rates.
The three coded schemes include the LT codes with specially optimized degree distribution (with or without a systematic phase), growth codes, and chunked codes. Particularly, with the LT codes, we are able to formulate the degree distribution design problem in the heterogeneous scenario into linear optimization problems. For the schemes compared, we characterize numerically the number of transmissions needed to deliver the demand of all users. The numerical evaluations agree with simulation results.

A systematic phase delivers the demands efficiently when the fraction of content
requested by the users does not exceed the link capacity.
On the other hand, for higher demands, coding significantly improves the delivery time. Different user demographics result in very different coding schemes being suitable for efficient delivery of demands.
Growth codes and chunked codes, are not found to be as suitable to the communication scenario as the optimized LT codes. Interestingly, time-shared broadcast of degraded message sets is found to
give a comparable delivery time performance as that of the non-systematic optimized LT codes.


As for future work, we are interested in incorporating the source coding stage into code design, since there is clearly an interplay between the compression efficiency in source coding and the efficiency of channel coding.
We are also interested in the exploration and analysis of more competitive schemes for broadcasting to heterogeneous users.

%% file: appendices.tex
\appendix
\section*{Proof of Claim \ref{thm:ft_randchannel}}
We give two independent arguments to validate the claim.

After the systematic phase, each content packet successfully reaches user $i$ independently at probability $1-\epsilon_i, $ and an average of $(1-\epsilon_i)N$ packets reaches the user. If $z_i\le 1-\epsilon_i,$ the demand of user $i$ is considered fulfilled. We consider the case where $z_i>1-\epsilon_i.$

In the first argument, we find out the non-systematic equivalent degree distribution $\bar P(x)$. To do that we only need to find out the (normalized) number of degree-1 packets(containing content packets selected uniformly at random with replacement) required to be transmitted in order for user $i$ to recover the $(1-\epsilon_i)N$ distinct content packets received in the systematic phase. This number, $-\frac{\ln\epsilon_i}{1-\epsilon_i},$ can be obtained by setting $P(x)=1\cdot x$(all-one degree distribution) in \eqref{eq:lt_find_ti}. A coupon collector's argument brings to the same conclusion\cite[Ch.~2]{feller} (see also \cite{monograph}):
 the expected number of samplings required to collect $zN$ distinct coupons is
\begin{align}
&N\Bigl(\frac{1}{N}+\frac{1}{N-1}+\dots+\frac{1}{N-zN+1}\Bigr)\gtrapprox N\ln\frac{N}{N-zN+1}=-N\ln\Bigl(1-\frac{zN-1}{N}\Bigr) \label{eq:coupon}.
\end{align}
Divide \eqref{eq:coupon} by $N(1-\epsilon_i)$, take $N\rightarrow\infty$ and let $z=1-\epsilon_i$, we find the same result as found from
\eqref{eq:lt_find_ti}.

Hence, $\bar p_1=-\frac{\ln\epsilon_i}{1-\epsilon_i}+p_1$ and $\bar P(x)=-\frac{\ln\epsilon_i}{1-\epsilon_i}x+P(x).$
In \eqref{eqn:rec_dec}, replace $P(x)$ by $\bar P(x)$ and $t_i$ by $(t_i-1)$ to subtract the time spent in the systematic phase, and we get \eqref{eq:rec_dec_sys}.

Alternatively, one can consider the packets received in the systematic phase as side information to the decoder, such as in \cite{dino,gummadi}.  These packets are removed from the subsequent coded packets, and we need to decode another $(z_i-(1-\epsilon_i))N$ packets from the remaining $\epsilon_iN$ packets. Let $d$ and $\hat d$ be the random variables representing the degree of a randomly generated coded packet and the degree of the coded packet after the removal of the packets received in the systematic phase. Thus, $\hat d=\sum_{j=1}^{d}X_j$ where $X_j(j=1,2,\dots,D)$ are i.i.d Bernoulli($\epsilon_i$) random variables indicating if the $j$th content packet participating in the coded packet has {\it not} been received in the systematic phase and remains in the coded packet. Thus, the moment generating function $\hat D$ is $\hat P(x)=P(Q(x))=P(1-\epsilon_i+\epsilon_ix)$ where $Q(x)=1-\epsilon_i+\epsilon_ix$ is the moment generating function of the i.i.d. $X_j$s. In \eqref{eqn:rec_dec}, replace $P(x)$ by $\hat P(x)$, $z_i$ by $\hat z_i=\frac{(z_i-(1-\epsilon_i))N}{\epsilon_iN}=\frac{z_i-(1-\epsilon_i)}{\epsilon_i}$, and $t_i$ by $\hat t_i=\frac{(t_i-1)N}{\epsilon_iN}=\frac{t_i-1}{\epsilon_i},$ we have
\begin{equation}
(1-\epsilon_i)\frac{t_i-1}{\epsilon_i}P'(Q(x))Q'(x)+\ln(1-x)>0, \quad \forall x\in\left[0,\frac{z_i-(1-\epsilon_i)}{\epsilon_i}\right).
\end{equation}
Let $y=Q(x)=1-\epsilon_i+\epsilon_ix,$ we get
\begin{equation}
(1-\epsilon_i)\epsilon_i\frac{t_i-1}{\epsilon_i}P'(y)+\ln\left(1-\frac{y-1+\epsilon_i}{\epsilon_i}\right)>0, \quad \forall y\in\left[1-\epsilon_i,1-\epsilon_i+\epsilon_i\frac{z_i-(1-\epsilon_i)}{\epsilon_i}\right),
\end{equation}
which is exactly \eqref{eq:rec_dec_sys}.